\documentclass[12pt]{article}
\usepackage{epsfig,cite}                                           
\def\beq{\begin{equation}}
\def\eeq{\end{equation}}
\def\eq{\beq\eeq}
\def\beqn{\begin{eqnarray}}
\def\eeqn{\end{eqnarray}}

\relax

\def\l{\left}

\def\eg{{\it e.g.}}
\def\ie{{\it i.e.}}

\def\pl#1#2#3{{\it Phys. Lett. } {\bf #1} (19#2) #3}

\def\pr#1#2#3{{\it Phys. Rev. } {\bf #1} (19#2) #3}
\def\np#1#2#3{{\it Nucl. Phys. } {\bf #1} (19#2) #3}

\def    \hepph  #1 {{\tt hep-ph/#1}}
\def    \hepex  #1 {{\tt hep-ex/#1}}
\def\eq#1{eq.~(\ref{#1})}

\catcode`@=11 
\def\gsim{\mathrel{\rlap{\lower4pt\hbox{\hskip1pt$\sim$}}
    \raise1pt\hbox{$>$}}}         
\def\lsim{\mathrel{\mathpalette\@versim<}}
\def\@versim#1#2{\lower0.2ex\vbox{\baselineskip\z@skip\lineskip\z@skip
       \lineskiplimit\z@\ialign{$\m@th#1\hfil##$\crcr#2\crcr\sim\crcr}}}
\catcode`@=12 
\def\epm#1#2{\hbox{${\lower1pt\hbox{$\scriptstyle +~#1$}}
\atop {\raise1pt\hbox{$\scriptstyle -~#2$}}$}}

\newcommand{\secn}[1]{section~\ref{#1}}

\newcommand\sss{\scriptscriptstyle}

\newcommand\as{\alpha_{\sss S}}

\def\blan{\Big\langle}   
\def\bran{\Big\rangle} 
\def\bea{\begin{eqnarray}}
\def\eea{\end{eqnarray}}

\def\l{\left}
\def\r{\right}

\begin{document}

\begin{flushright}
 
{\tt hep-ph/0205286} 
\\ RM3--TH/02-07 \\ DFTT 13/02 \\ GeF/TH/4--02 \\ UB--ECM--FP--02--12

\end{flushright}
 
\begin{center}

\vspace*{.6cm}

{\Large \bf Determination of $\as$ 
from scaling violations \\ of truncated moments of structure functions}

\vspace*{1.3cm}

{\bf Stefano Forte,$^{a}$
Jos\'e I. Latorre,$^{b}$
Lorenzo Magnea$^{c}$ and 
Andrea Piccione$^{d}$}

\vspace{0.5cm}

{\small
{}$^a$INFN, Sezione di Roma Tre\\
Via della Vasca Navale 84, I-00146 Rome, Italy\\

\vspace{0.2cm}

{}$^b$Departament d'Estructura i Costituents de la Mat\`eria, \\
Universitat de Barcelona, Diagonal 647, E-08028 Barcelona, Spain

\vspace{0.2cm}

{}$^c$Dipartimento di Fisica Teorica,
Universit\`a di Torino and INFN, Sezione di Torino\\
Via P.~Giuria 1, I-10125 Torino, Italy\\

\vspace{0.2cm}

{}$^d$Dipartimento di Fisica, Universit\`a di Genova and INFN,
Sezione di Genova,\\
Via Dodecaneso 33, I-16146 Genova, Italy\\
}

\vspace*{2.5cm}
                                                                 
{\bf Abstract}

\end{center}
\noindent

We determine the strong coupling $\as (M_Z)$ from scaling violations
of truncated moments of the nonsinglet deep inelastic structure
function $F_2$. Truncated moments are determined from BCDMS and NMC data
using a neural network
parametrization  which retains the full
experimental information on errors and correlations. Our method
minimizes all sources of theoretical uncertainty and bias which
characterize extractions of $\alpha_s$ from scaling violations.  We
obtain $\as(M_Z) = 0.124 ~\epm{0.004}{0.007} ~\hbox{(exp.)}
~\epm{0.003}{0.004} {}~\hbox{(th.)}$.


\vfill 
\begin{flushleft} 
May 2002 
\end{flushleft}

\eject   

\section{Introduction}
\label{intro}

As the predictions of QCD become increasingly precise and more high
quality data are available, the theoretical uncertainties associated
with the analysis of the data are more and more often found to be
dominant, and have thus come under increased scrutiny~\cite{qcd}.  The
determination of the QCD coupling $\alpha_s(M_Z)$~\cite{sb} from
scaling violations of deep inelastic structure functions is perhaps
the simplest and most fundamental example of this situation.  In
principle, deep inelastic scattering would be expected to provide a
theoretically solid and experimentally clean way of determining
$\alpha_s(M_Z)$.  In practice, however, the situation is far from
satisfactory~\cite{altarev}, as revealed by the lack of stability of
the value of $\alpha_s$ obtained through this procedure.

The main source of difficulties can be traced to the fact that
conventional extractions of $\alpha_s$ from scaling violations of
structure functions involve a simultaneous determination of parton
distributions of the nucleon. The error on $\alpha_s$, therefore, gets
tangled with the uncertainty on parton distributions, which is
notoriously difficult to assess, and subject to a variety of sources
of theoretical bias~\cite{qcd}.  This is a consequence of the fact
that the evolution equations of perturbative QCD~\cite{AP}, which
govern scaling violations, can only be solved analytically in compact
form by taking Mellin moments, which diagonalize the evolution
operator.  In practice, however, Mellin moments are not directly
measurable, since their determination would require the availability
of data taken at arbitrarily large center-of-mass energy.  The
problem is circumvented by solving directly the momentum-space
Altarelli-Parisi~\cite{esw} equations, which, however, usually entails the
construction of a parton parametrization.

Such an undertaking thus runs into two difficulties.  First, a parton
parametrization is routinely constructed by assuming a functional form
for parton distributions, and fitting the corresponding free
parameters.  The choice of a functional form, however, introduces a
theoretical bias, which is potentially large, and whose size is very
difficult to assess~\cite{pdfrev}. For example, choices of functional
form based on ``QCD expectations'' may well give a
poor representation of unexpected or unconventional behaviors of the
distributions. Furthermore, any given parametrization constrains the
form of the distribution near and beyond the boundary of the region
where data are available: whenever the data are not very precise, it
can be seen~\cite{abfr} that the results obtained for observable
quantities may depend significantly on the form of the
parametrization.

Second, it is quite difficult to estimate precisely the error on
parton distributions, and even harder to propagate this error to
quantities which depend upon them. Indeed, a generic observable (such
as a Mellin moment, or a cross section) is a functional of parton
distributions, which in general depends on it in a complicated,
nonlocal way.  Thus, on the one hand it is difficult to extract the
errors on parton distributions from errors on the data, on the other
hand it is difficult to propagate errors on parton distributions to
quantities calculated from them. At the very least, the full
information on experimental errors and correlations should be used,
but it may also turn out that conventional techniques of linear error
propagation fail.  These difficulties could only be overcome if one
were able to obtain a reliable and unbiased representation of the
probability measure in the functional space of parton
distributions~\cite{pdfer}, as determined by the available
experimental information.

In this paper, we approach the determination of $\alpha_s$ in a way
which bypasses these difficulties, by combining two techniques which
have recently been proposed and implemented in the context of the
analysis of DIS data: the use of evolution equations for truncated
Mellin moments of parton distributions~\cite{us1,us3,deer}, and the
determination of the probability measure in the space of deep
inelastic structure functions by means of neural networks~\cite{us2}.

Truncated moments are defined as Mellin moments with the integration
range restricted to the subset $x_0 < x < 1$ of the kinematically
allowed range $0 < x < 1$. Unlike ordinary Mellin moments, they are
measurable quantities, since the small $x$ region is excluded.  The
evolution equations which they satisfy turn out to be a system of
coupled ordinary linear differential equations, which can be truncated
to finite accuracy to yield a closed system.  The evolution equations
for truncated moments have been derived, and explicitly solved at
next-to-leading order in the nonsinglet~\cite{us1} and
singlet~\cite{us3} sectors. The numerical accuracy of the approximate
solutions has been studied and improved upon by means of suitable
techniques~\cite{deer}, and is equal or better than that of standard
numerical solutions of Altarelli-Parisi equations~\cite{qcd}.

Given an experimental determination of truncated moments at more than
one scale, one can then determine $\alpha_s$ directly by comparing 
computed and measured scaling violations. In the nonsinglet sector,
there is only one independent structure function, say $F_2$, and
scaling violations are fully determined by $\alpha_s$ and an initial
condition, given by the values of the truncated moments of $F_2$ at a
reference scale: therefore one does not have to use a parton
parametrization.

In practice, however, a sufficiently accurate determination of
truncated moments of the nonsinglet $F_2$ and associated errors and
correlations cannot be obtained by simply binning and summing data
points: the coverage and precision of the data are not sufficient for
this purpose.  Rather, in order to fully exploit the available
experimental information, it is necessary to construct a
parametrization of the structure function in the kinematic region
where data are available.  This clearly leads back to the same
difficulties encountered when constructing a parton
parametrization. The problem, however, can now be overcome, since an
unbiased determination of the probability density in the space of
structure functions, based on available experimental data, has been
constructed in Ref.~\cite{us2}.

The representation of the probability density given in Ref.~\cite{us2}
takes the form of a set of neural networks, trained on an ensemble of
Monte Carlo replicas of the experimental data, which reproduce their
probability distribution.  The parametrization is unbiased, since
neural networks do not require the choice of a specific functional
form, and it interpolates between data points, imposing smoothness
constraints in a controllable way.  Information on experimental errors
and correlations is incorporated in the Monte Carlo sample, so that
errors on truncated moments and correlations between them can be
determined without having to use linearized error propagation. The
parametrization combines all the available experimental information,
so that statistical errors are minimized, but it also correctly
reproduces the loss of accuracy incurred when extrapolating outside
the kinematical region where data are available.

Hence, we can obtain an unbiased evaluation of truncated moments, and
then use this to determine $\alpha_s$ directly without any further
assumptions. We end up with a determination of $\alpha_s$ where all
sources of uncertainty related to the method of analysis are under
control, and the only theoretical uncertainty is related to the lack
of knowledge of the anomalous dimensions beyond next-to-leading
order. This gives us a bias-free determination of $\alpha_s$, and
simultaneously illustrates the power of a method of analysis based on
the direct knowledge of a probability density in a space of functions.

This paper is organized as follows: in \secn{trumo} we will review the
method of truncated moments~\cite{us1,us3,deer}, emphasizing the
issues of numerical accuracy which are relevant to the rest of the
paper; in \secn{neune} we will introduce the neural network
parametrization of DIS structure functions developed in
Ref.~\cite{us2}; \secn{detal} contains the details of our fitting
procedure and our result for the strong coupling $\as(M_Z)$: we
explain our data selection, our choice of fitting architecture, our
error estimate and the consistency tests that we performed; finally,
\secn{conre} summarizes our conclusions and discusses possible
extensions of our work.

\section{Truncated moments of parton distributions}
\label{trumo}

We determine the strong coupling from the scale dependence of the
nonsinglet deep inelastic structure function
\beq
F_2^{NS} (x, Q^2) = F_2^{p} (x, Q^2) - F_2^{d} (x, Q^2)~.
\label{f2ns}
\eeq
In the DIS scheme~\cite{DIS}, $F_2^{NS}$ is expressed directly in
terms of the nonsinglet combination of quark distribution, according
to
\beq
F_2^{NS} (x, Q^2) = \sum_{i = 1}^{n_f} e_i^2 \left[ q_i (x, Q^2) +
\overline{q}_i (x, Q^2) \right]_{p - n}~.
\label{f2ns2}
\eeq
The scale dependence of $F_2$ is thus given by the evolution equation
for nonsinglet quark distributions, henceforth denoted simply by $q(x,
\mu^2)$,
\beq
\mu^2 \frac{d}{d \mu^2}~q(x, \mu^2) =
\frac{\as (\mu^2)}{2 \pi} \int_x^1 \frac{d y}{y} 
P \left(\frac{x}{y}, \as(\mu^2) \right) q(y, \mu^2)~,
\label{alpar}
\eeq
where $P(z, \as)$ is the appropriate  evolution kernel.  

\subsection{Evolution equations for truncated moments}
\label{eetm}

Consider now the truncated Mellin moment of the quark distribution
\beq
q_n (x_0, \mu^2) \equiv \int_{x_0}^1 d x ~x^{n - 1} q(x, \mu^2)~.
\label{trunc}
\eeq
The evolution equation for truncated moments is easily found to be
\beq
\mu^2 \frac{d}{d \mu^2} ~q_n(x_0, \mu^2) = \frac{\as 
(\mu^2)}{2 \pi} \int_{x_0}^1 d y ~y^{n - 1} q(y, \mu^2) ~G_n 
\left(\frac{x_0}{y}; \as(\mu^2)\right)~,
\label{truncalpar}
\eeq
where $G_n$ is the truncated moment of the splitting function
\beq
G_n(x, \as) = \int_x^1 d z z^{n - 1} P(z, \as)~.
\label{kern}
\eeq
Note that the function $G_n (x, \as)$ is analytic on the real axis in
the interval $0 \leq x < 1$, while it has integrable singularities at
$x = 1$ as a consequence of the presence of $+$ distributions in the
kernel $P(z, \as)$ at all perturbative orders.

We can turn \eq{truncalpar} into a set of coupled linear differential
equations by expanding $G(x_0/y, \as)$ in powers of $y$ around $y = 1$
as
\beq
G_n \left(\frac{x_0}{y}, \as \right) = \sum_{p=0}^\infty ~\frac{1}{
p!} ~g_p^{(n)} (x_0, \as) ~(y - 1)^p~.
\label{ser1}
\eeq
Because the singularities of $G_n (x_0/y, \as)$ at $y \to x_0$ are
integrable, one can substitute \eq{ser1} into \eq{truncalpar} and
integrate term by term, obtaining a convergent series. It is then
legitimate to truncate the sum in \eq{ser1} at a finite order $p = M$,
with the result
\beq
G_n \left(\frac{x_0}{y}, \as \right) = \sum_{p=0}^{M} ~c_{p,n}^{(M)} 
(x_0, \as) ~y^p + O\left[(y - 1)^{M + 1} \right]~,
\label{ser2}
\eeq
where
\beq
c_{p,n}^{(M)}(x_0, \as) = \sum_{k=p}^{M} 
\frac{(-1)^{p + k} g_k^{(n)} (x_0, \as)}{p! (k - p)!}~.
\label{co2}
\eeq
The evolution equation for truncated moments then becomes~\cite{us1} 
\beq
\mu^2 \frac{d}{d \mu^2} q_n (x_0, \mu^2) =
\frac{\as (\mu^2)}{2 \pi} \sum_{p=0}^{M} ~c_{p,n}^{(M)}(x_0, 
\as) ~q_{n + p}(x_0, \mu^2)~,
\label{finsyst}
\eeq
and is thus  given by a set of coupled linear differential
equations. The anomalous dimension matrix $c_{p,n}^{(M)}(x_0, \as)$
couples each truncated moment $q_n$ to all truncated moments $q_k$
with $k \ge n$. The evolution equation for the $k$-th moment can be
made arbitrarily accurate by taking $M$ to be sufficiently large.

\subsection{Solution of the evolution equations}
\label{see}

The coupled evolution of the moments $q_k$, with $n \leq k \leq n +
M$, can be solved to determine the value of $q_n$ at different scales,
if the system in \eq{finsyst} closes, \ie\ if it is legitimate to
include a decreasing number of terms in the evolution equation of
increasingly higher moments.  In Ref.~\cite{us1} it was shown that
this is indeed the case, because the matrix elements of $c_{p,n}^{(M)}
(x_0, \as)$ decrease rapidly as one moves away from the diagonal, so
that the accuracy of the truncated evolution kernel actually improves
with increasing order of the moment, even if one less term is included
when the order is increased by one unit.

Having chosen the values for $n$ and $M$, we can introduce the
simplified notation
\beq
\left.
\begin{array}{cccc}
C_{k l}(\as) & = & c_{l - k,k}^{(M - k + n)} (x_0, \as) & (n + M \geq l 
\geq k \geq n)~, \\
C_{k l}(\as) & = & \hspace{-2cm} 0 & (n \leq l < k \leq n + M)~,
\end{array}
\right.
\label{matr0}
\eeq
in terms of which the system of equations to be solved reads simply
\beq
\mu^2 \frac{d}{d \mu^2} q_k (x_0, \mu^2) = \frac{\as (\mu^2)}{2 \pi}
\sum_{l = n}^{n + M} C_{k l}(\as) ~q_l (x_0, \mu^2)~,
\label{simp}
\eeq
with $n \leq k \leq n + M$. Expanding $C_{k l}(\as)$ in powers of
$\as$ to next-to-leading order, one can now solve \eq{simp} by
standard perturbative methods. The task is considerably simplified by
the fact that the matrix of coefficients, $C_{k l}(\as)$, is upper
triangular, so that the eigenvalues are exactly the diagonal elements
$C_{k k}$, and the eigenvectors are easily computed by means of a
recursion relation.  The explicit solution is given in
Ref.~\cite{us1}.

In practice, the numerical implementation of the solution may be
problematic if a high accuracy is required for low moments.  Indeed,
it turns out that when $x_0 = 0.1$ the value of $M$ which is needed to
achieve an accuracy at the few percent level on the evolution of the
$n$-th truncated moment is typically of order $M \gsim 100$ for the
lowest moments, $n = 1$, $2$ (when $n < 1$ the convolution integral in
\eq{truncalpar} diverges), even though it rapidly decreases for higher
moments. Such a large value of $M$ leads to numerical difficulties,
due to the fact that the matrix elements $C_{k l}(\as)$ become very
small when $l >\!> k$.

\subsection{Improved solution}
\label{impso}

An improved solution to the evolution equation, which overcomes the
problems related to the slow convergence of the expansion of the
evolution kernel for low moments, can however be
derived~\cite{deer}. The origin of the problem is easily tracked to
the logarithmic singularities left over in the function $G_n(x_0/y,
\as)$, for $y \to x_0$, as a consequence of the presence of $+$
distributions in the evolution kernel $P(z, \as)$, for $z \to 1$.  For
the lowest moments, the integration over $y$ is dominated by small
values of $y$, $y \sim x_0$, where the function $G_n$ is in turn
dominated by
the logarithmic singularity, and not well approximated by  a low-order
truncation of its Taylor
expansion around $y = 1$.

To solve the problem, one can integrate by parts the right hand side
of \eq{truncalpar}, obtaining
\bea
&& \int_{x_0}^1 d y ~y^{n-1} ~q(y, \mu^2) ~G_n \left( \frac{x_0}{y}, \as 
\right) = \left[\widetilde{G}_n (x_0, y; \as) ~y^{n-1} ~q(y,Q^2) 
\right]_{x_0}^1 \nonumber \\ 
&& \hspace{5mm} - ~\int_{x_0}^1 dy ~\widetilde{G}_n (x_0, y; \as) 
\frac{d}{dy} \left(y^{n-1} q(y,Q^2) \right)~, 
\label{rhs} 
\eea
where
\beq
\widetilde{G}_n (x_0, y; \as) = \int_{x_0}^y dz ~G_n \left(
\frac{x_0}{z}, \as \right)~.
\label{gt1}
\eeq
In the definition of $\widetilde{G}_n$ the freedom to fix the lower
limit of integration, which is independent of $y$, has been exploited,
so that the function $\widetilde{G}_n$ satisfies $\widetilde{G}_n
(x_0, x_0; \as) = 0$, as well as
\beq
\widetilde{G}_n (x_0, y; \as) = y ~G_n \left( \frac{x_0}{y}, \as \right) - 
x_0 ~G_{n-1} \left(\frac{x_0}{y}, \as \right)~;
\eeq
furthermore, the coefficients of the Taylor expansion of
$\widetilde{G}_n$ around $y = 1$, defined as in \eq{ser1}, satisfy, in
obvious notations,
\beq
\widetilde{g}_p^{(n)} (x_0, \as) = g_{p - 1}^{(n)} (x_0, \as)~.
\label{dertaycoeff}
\eeq
The integration region $y \sim x_0$ is now suppressed, and 
a faster rate of convergence when $\widetilde{G}_n$ is expanded around
$y = 1$ is expected. 
Repeating the procedure that leads  to \eq{simp} one gets~\cite{deer}
\bea
\mu^2 \frac{d}{d \mu^2} q_n (x_0, \mu^2) & = & \frac{\as(\mu^2)}{2 \pi} 
\left[ x_0^{n - 1} q(x_0,\mu^2) \sum_{p = 0}^{M} 
\frac{1}{p!} ~\widetilde{g}_p^{(n)} (x_0, \as) (x_0 - 1)^p 
\right. \nonumber \\ & + & \left. \sum_{l = 1}^{M} 
C_{nl}(\as) q_{l} (x_0, \mu^2) \right]~.
\label{eqev1}
\eea
The coefficients $C_{nl}(\as)$ are given in \eq{matr0}. The first term
on the right hand side of \eq{eqev1} is responsible for the faster
convergence of the expansion to the exact result: it vanishes as $M
\to \infty$, as it must, since it reconstructs $\widetilde{G}_n (x_0,
x_0; \as) = 0$.

Eq.~(\ref{eqev1}), however, cannot be used directly: as
the right-hand side of the equation depends on the value of $q(x_0,
\mu^2)$, we do not get a closed system of equations. This obstacle can
be circumvented by noting~\cite{deer} that $q(x_0, \mu^2)$ itself can
be expanded over a basis of truncated moments, and, specifically, a
finite set of them is sufficient to give an accurate determination of
$q(x_0, \mu^2)$.  To see this, expand the quark distribution in Taylor
series about a selected point, say  $ y_0 = (1 + x_0)/2$ (note
that $y_0 > (1 + x_0)/2$ is not allowed since $y = 1$ is an essential
singularity of all parton distributions)
\beq
q(x, \mu^2) = \sum_{k = 1}^{\infty} \eta_k (\mu^2) ~(x - y_0)^{k - 1}~.
\label{pdftaylor}
\eeq
Truncated moments are then given by
\beq
q_j (x_0, \mu^2) = \sum_{k = 1}^{\infty} \beta_{j k} ~\eta_k(\mu^2)~,
\label{qjx0}
\eeq
where
\beq
\beta_{j k} = \int_{x_0}^1 dx ~x^{j-1} (x - y_0)^{k-1}~.
\label{betadef}
\eeq

The infinite matrix $\beta_{j k}$ can be approximated by truncating it
to a square $N \times N$ matrix $\widetilde{\beta}_{jk}$, which is
easily computed and inverted. One finds then an approximate expression
for $q(x_0, \mu^2)$, given by
\beq
q(x_0, \mu^2) =
\sum_{k = 1}^{N} \sum_{j = 1}^{N} \widetilde{\beta}^{-1}_{kj} \,
q_j (x_0, \mu^2) \, (x_0 - y_0)^{k - 1}~.
\label{pdfapprox}
\eeq
The error introduced by the truncation of the matrix $\beta_{j k}$ has
been studied, as a function of $N$, in Ref.~\cite{deer}. It turns out
that a satisfactory accuracy (at the $0.1 \%$ level) can be reached
already for $N \sim 5$, while in practice the inversion of the matrix
$\widetilde{\beta}_{jk}$ becomes numerically difficult only for $N >
10$ (as easily verified, the matrix is ill-conditioned).  The method
is thus viable and does not lead to loss of accuracy.

Combining \eq{eqev1} and \eq{pdfapprox}, we get the final form for the
evolution equation, which replaces \eq{simp}, namely
\beq
\mu^2 \frac{d}{d \mu^2} q_k (x_0, \mu^2) = \frac{\as (\mu^2)}{2 
\pi} ~\sum_{l = n}^{n + M} \left[C_{k l}(\as) + D_{k l}^{(N)} (\as) 
\right]~q_l (x_0, \mu^2)~,
\label{simp2}
\eeq
where
\beq
D_{k l}^{(N)} (\as) = x_0^{k - 1} \left[ \sum_{p = 0}^{M} 
\frac{1}{p!} ~\widetilde{g}_p^{(k)} (x_0, \as) (x_0 - 1)^p \right]   
\left[\sum_{r = 1}^N \widetilde{\beta}^{-1}_{r l} ~(x_0 - 
y_0)^{r - 1} \right]~.
\label{Dnl}  
\eeq

Once again, \eq{simp2} can be solved by expanding $C_{k l}(\as)$ and
$D_{k l}^{(N)}(\as)$ to next-to-leading order in $\as$, and then
using well-known perturbative methods. A compact expression for the
solution is given in Refs.~\cite{us3,deer}. The price to pay for the
faster convergence of \eq{simp2} is that the evolution matrices are no
longer triangular. This implies that all truncated moments with $n \ge
1$ must be used as initial conditions in order to compute the
evolution of any moment.  As a consequence, diagonalization has to be
performed numerically rather then analytically.  In practice, however,
this does not lead to any numerical difficulties: as shown in
Ref.~\cite{deer}, an accuracy at the few percent level on the
evolution can be achieved for any truncated moment with $M \sim 10$, a
size at which numerical diagonalization is not difficult.  It is
important to notice that all the approximations that have been
introduced in order to make the problem amenable to a numerical
solution are under theoretical control, and the accuracy achieved can
be precisely estimated and can be improved upon if the need arises.

\section{\hspace{-6pt}Neural network parametrization of deep inelastic 
structure functions}
\label{neune}

Ideally, a parametrization of structure functions must incorporate all
the information contained in the experimental measurements, {\it i.e.}
their
central values, their statistical and systematic errors, and their
correlations; furthermore, it must interpolate between them without
introducing any bias. Several approaches to this problem have been
proposed, in the context of fitting parton distributions:
$\chi^2$-minimization of a fixed functional form coupled to error
propagation through the covariance matrix, and various improvements
thereof~\cite{ale}; projection over bases of orthogonal
polynomials~\cite{ortho}; or Monte Carlo sampling coupled to Bayesian
inference~\cite{pdfer}.  Here we will follow the method of
Ref.~\cite{us2}, where an unbiased extraction of the probability
measure in the space of structure functions is performed, based on a
coordinated use of Monte Carlo generation of data and neural network
fits.

\subsection{Experimental data}
\label{expda}

\begin{figure}[t]
\begin{center}
\epsfig{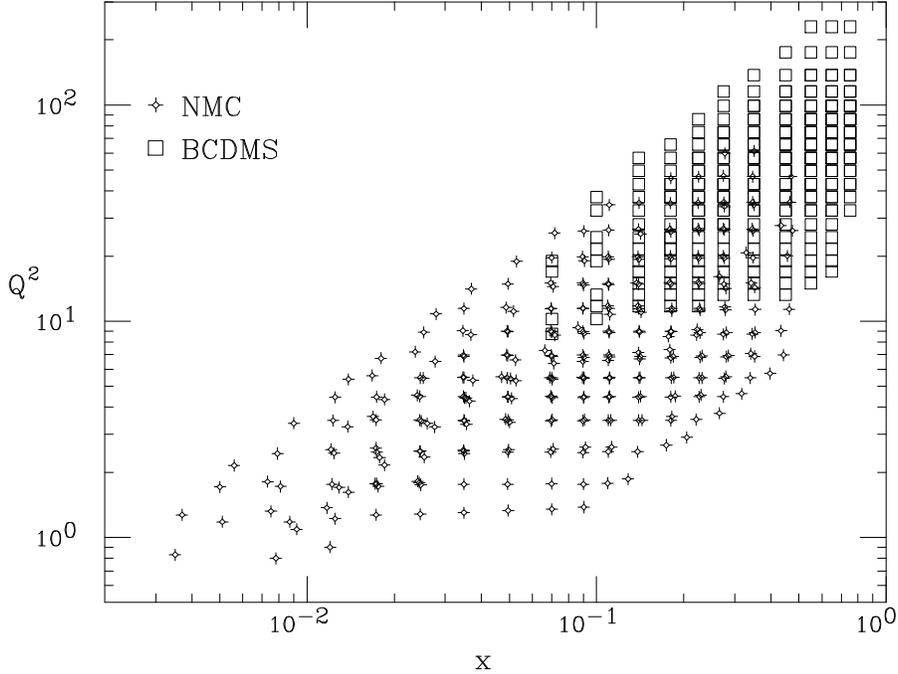}
\end{center}
\begin{center}
\caption{NMC and BCDMS kinematic range.}
\label{fig:kinrange}
\end{center}
\end{figure}

Since we are interested in the nonsinglet structure function $F_2$, we
need a simultaneous measurement  of this structure
function for proton and deuterium targets.  We will use the data of
the New Muon Collaboration (NMC)~\cite{NMC} and of the BCDMS
(Bologna-CERN-Dubna-Munich-Saclay) Collaboration~\cite{BCDMS},
which provide a simultaneous determination of the proton and deuteron
structure functions in the same kinematic region, and provide the full
set of correlated experimental systematics for these
measurements. Earlier data from SLAC are not competitive with these in
terms of accuracy and kinematic coverage. The more recent HERA data
are available in a much wider kinematic region, but only for proton
targets. Another set of joint proton and deuteron measurements was
performed by the E665 Collaboration \cite{e665}. These data however
are mostly concentrated at low $Q^2$ (and low $x$), and thus are not
relevant for perturbative QCD applications. The kinematic coverage of
the data which we include in our analysis is displayed in
Fig.~\ref{fig:kinrange}.
Altogether our parametrization is based on $N_{dat} = 552$
experimental points for the nonsinglet structure function, obtained as
difference of pairs of proton and deuteron data according
to~\eq{f2ns}.
Henceforth, $F_i^{(exp)}$ will denote the $i$-th data point $F_2^{NS}(x_i,Q^2)$.

\subsection{Probability measure in the space of structure functions}
\label{proda}

The experimental data give us a probability measure in an
$N_{dat}$-dimensional space, assumed to be multigaussian.  In order
to extract from it a parametrization of the desired structure function
we must turn this measure into a measure ${\cal P}[F_2]$ in a space of
functions.  Once such a measure is constructed, the expectation value
of any observable ${\cal F} \left[F_2(x, Q^2)\right]$ can be found by
computing the weighted average
\beq
\blan {\cal F} \left[ F_2(x,Q^2) \right] \bran = \int \! 
{\cal D} F_2 \, {\cal F} \left[ F_2(x,Q^2) \right] \, {\cal P}[F_2]~.
\label{funave}
\eeq
Errors and their correlations can also be obtained from this
measure, by considering higher moments of the same observable with
respect to the probability distribution. 

The determination of an infinite-dimensional measure from a finite
set of data points is an ill-posed problem, unless one introduces
further assumptions.  For instance, one may assume a fixed functional
form~\cite{ale}, in which case the problem is reduced to the
determination of a finite set of parameters, or one may project over a
basis of orthogonal polynomials~\cite{ortho}, and choose a particular
truncation of the {\it a priori} infinite set of basis functions. In
the approach of Ref.~\cite{us2}, neural networks are used as
interpolating functions, so that the only  assumption is the
smoothness of the structure function. Neural networks can fit
any continuous function through a suitable training; smoother
functions require a shorter training and less complex networks.
Hence, an ideal degree of smoothness can be established on the basis
of a purely statistical criterion (such as having reached a
satisfactory goodness-of-fit) without need for further assumptions.

\subsection{Fitting strategy}
\label{pseda}

The construction of the probability measure is done in two steps:
first, a set of Monte Carlo replicas of the original data set is
generated. This gives a representation of the probability density
${\cal P}[F_2]$ at points $(x_i, Q^2_i)$ where data are
available. Then, a neural network is fitted to each replica.  The
ensemble of neural networks gives a representation of the probability
density for all $(x,Q^2)$: when 
interpolating between data the uncertainty  will be kept under control
by the smoothness constraint, but it will become increasingly more
sizable when extrapolating away from the data region.

The $k = 1,\dots, N_{rep}$ replicas of the data are generated as
\beq
F_i^{(art)\,(k)} = (1 + ~r_{i,N}^{(k)} ~\sigma_{i,N}) \Bigg[ F_i^{(exp)}
+ \frac{ \sum_a r_{i,a}^{(k)} f_{i,a}}{100} ~F_i^{(exp)} +
r_{i,s} ~\sigma_{i,s}^{(k)} \Bigg]~,
\label{gennmc}
\eeq
where $F_i^{(exp)} \equiv F_2(x_i, Q^2_i)$ are the original data,
$f_{i,a}$ are the experimental systematic errors, given in percentage,
$\sigma_{i,N}$ is the experimental normalization error, $\sigma_{i,s}$
is the experimental statistical error, while $r_{i,a}^{(k)}$,
$r_{i,s}^{(k)}$ and $r_{i,N}^{(k)}$ are univariate gaussian
numbers. When two data points share a correlated error or
normalization, the same gaussian number is used. In Ref.~\cite{us2}, a
set of $N_{rep} = 1000$
replicas of this form has been  generated, and it has been verified that
central values, errors and correlations of the original experimental
data are well reproduced by taking the relevant averages over a sample
of this size. The kinematic bound $F_2(1,Q^2)=0$ has been
implemented by adding  a number of artificial data points at $x=1$
with carefully adjusted errors.

Each set of generated data is fitted by an individual neural
network. A neural network~\cite{netrev} is a function of a number
of parameters, which fix the strength of the coupling between neurons
and the thresholds of activation for each neuron. The architecture of
the network is chosen to be redundant for the given problem, \ie\ the
number of parameters is rather larger than the minimum needed to get a
good fit.  The parameters are then fitted by minimizing an error
function. The fit is done using the technique of learning by
back-propagation, whereby the data are repeatedly shown to the
network until satisfactory learning is achieved.  The error function
is
\beq
E^{(k)} = \sum_{i = 1}^{N_{dat}} \frac{ \left( F^{(art)(k)}_i - 
F^{(net)(k)}_i \right)^2}{{\sigma^{(exp)}_{s,i}}^2},
\label{diaenergy}
\eeq
where $F^{(net)(k)}_i$ is the prediction for the $i$-th data point
from the net trained on the $k$-th replica of the data.  Use of this
error functions, which only includes statistical errors,
ensures~\cite{us2} that statistical errors on adjacent data points are
correctly combined, whereas the correlated systematics are reproduced
when averaging over the Monte Carlo sample.

\subsection{Results and validation}
\label{valda}

In Ref.~\cite{us2}, an independent set of neural networks was trained
on the non-singlet combination $F_i^{(p)} - F_i^{(d)}$.  This
procedure is advisable because the structure functions are roughly of
the same size, while their difference is typically by a factor 10
smaller.  The length of training is fixed by studying the behaviour of
the error function $E^{(0)}$, defined as in~\eq{diaenergy} for the
neural net fitted to the central experimental values, and asking that
$E^{(0)}/N_{dat}$ stabilizes to a value close to one.

A number of checks is then performed in order to make sure that an
unbiased representation of the probability density has been
obtained. For instance, it has been verified that 
the covariance of two data points computed
from the Monte Carlo sample of nets is on average very close to the
corresponding covariance matrix element of the data. Since
correlations of the data are entirely due to systematics, this
indicates that the systematics is correctly reproduced.

It turns out that the average standard deviation for each data point
computed from the Monte Carlo sample of nets is substantially smaller
than the experimental error. This could be due to the fact that the
network is combining the information from different data points by
capturing an underlying law, or that it is introducing a smoothing
bias. In  Ref.~\cite{us2} a statistical indicator which
distinguishes the two cases was constructed: define
\beq
\widetilde{E}^{(k)} = 
\sum_{i = 1}^{N_{dat}} \frac{\l( F^{(exp)(k)}_i - F^{(net)(k)}_i \r)^2}
{{\sigma^{(exp)}_{i,s}}^2}~,
\label{modavendef}
\eeq
{\it i.e.} a modified error function where the prediction of the $k$-th
net for the $i$-th point is compared to the central experimental
value rather than to the $k$-th replica.  The desired indicator is
\beq
{\cal R}\equiv \frac{\langle \widetilde{E} \rangle_{rep} }{\langle  
E\rangle_{rep}}~,
\label{calrdef}
\eeq
where $\langle \rangle_{rep}$ denotes the average over the ensemble of
replicas, and $E$, $\widetilde{E}$ are respectively defined by
eqs.~(\ref{diaenergy}) and (\ref{modavendef}).  One can then show
that, in the limit in which the error computed from the Monte Carlo
sample is much smaller than the experimental one, ${\cal R} \approx
1/2$ if an underlying law is captured by the net, but ${\cal R} \gsim
1$ if the error reduction is due to a smoothing bias. The Monte Carlo
sample of nets of Ref.~\cite{us2} has ${\cal R} = 0.58$.
 
The final set of neural networks $ F_i^{(net)(k)} $ provides a
representation of the probability measure in the space of structure
functions, which can be used to estimate any functional average,
defined as in~\eq{funave}, using
\beq
\blan {\cal F}
\left[F_2(x,Q^2)\right] \bran = \frac{1}{N_{rep}} 
\sum_{k = 1}^{N_{rep}} {\cal F} \left[{F_2}^{(net)(k)} (x,Q^2)
\right]~.
\label{discfunave}
\eeq
In particular, the average, variance, and covariance of truncated
moments computed using the Monte Carlo sample will give us a
determination of values, errors and correlations of the measured
truncated moments.

\section{Determination of $\as$}
\label{detal}

Using the neural parametrization of structure functions, we can
compute directly experimental values of  truncated moments at any
scale in the region of the data.  Because the neural parametrization
retains all the experimental information, and specifically it allows a
determination of errors and correlations, we can view such values as
direct experimental determinations of the truncated moments.  The
value of $\as$ can then be extracted from scaling violations of
truncated moments.  Specifically, given the full set of truncated
moments at a reference scale, the value of any truncated moment at any
other scale is predicted in terms of $\as$. Hence, we can obtain a
determination of $\as$ by comparing this prediction with the data, for
any choice of moment and scale. Even though such predictions are
clearly correlated, it is to be expected that the use of a larger
number of moments or scales will in general lead to a more precise
determination of $\as$; the accuracy should, however, saturate for a
large enough number of moments and scales.  Errors and correlations
for typical truncated moments are displayed in Tables~\ref{cormom} and
\ref{corscal}. Correlations are quite large, so it is conceivable that
an optimal fit may be obtained with a relatively small number of
moments.
\begin{table}[t]  
\begin{center}  
\begin{tabular}{|l|ccccc||c|}\hline 
n & 2      & 3      & 4      & 5      &  6      &  $\sigma$ (\%) \\
\hline 
2 & 1.0    & 0.966  & 0.895  & 0.808  &  0.718  &  8.8     \\
3 & 0.966  & 1.0    & 0.977  & 0.923  &  0.854  &  7.5     \\
4 & 0.895  & 0.977  & 1.0    & 0.983  &  0.941  &  7.4     \\
5 & 0.808  & 0.923  & 0.983  & 1.0    &  0.987  &  8.0     \\
6 & 0.718  & 0.854  & 0.941  & 0.987  &  1.0    &  8.9     \\
\hline
\end{tabular}
\end{center}
\caption{}{Errors and correlations for various truncated moments with
$x_0 = 0.03$ and $Q^2 = 20$~GeV$^2$.}
\label{cormom}
\end{table}
\begin{table}[t]  
\begin{center}  
\begin{tabular}{|l|ccccc||c|}\hline 
$Q^2 $   &  $20$  &  $27.4$ &  $37.4$  &  $51.2$  &  $70$   & $\sigma$ (\%) \\
\hline
$20  $   &  1.0   &  0.972  &  0.900   &  0.814   &  0.743  & 7.9  \\
$27.4$   &  0.972 &  1.0    &  0.977   &  0.926   &  0.870  & 7.2  \\
$37.4$   &  0.900 &  0.977  &  1.0     &  0.984   &  0.950  & 7.1  \\
$51.2$   &  0.814 &  0.926  &  0.984   &  1.0     &  0.988  & 7.3  \\
$70  $   &  0.743 &  0.870  &  0.950   &  0.988   &  1.0    & 7.3  \\
\hline
\end{tabular}
\end{center}
\caption{}{Errors and correlations for the fourth truncated moment with
$x_0 = 0.03$ at various scales  $20 \leq Q^2 \leq 70$~GeV$^2$.}
\label{corscal}
\end{table}

It is clear that to achieve a best fit for the strong coupling we must
optimize the choice of the fitting procedure and parameters. We now
turn to a detailed explanation of our choices for the truncation point
of Mellin moments, for the range and number of scales $Q^2$, for the
number of moments to be used in the approximate evolution equation,
the number of moments that should be treated as free parameters, and
the effect of variations of these choices. We will then give our best
fit with its associated statistical error. Finally, we will address
the known sources of theoretical  error.

\subsection{Choice of truncation point and fitting range}
\label{chora}

The choice of values of $(x_0$, $Q^2)$ and of the moments to be used
for the extraction of $\as$ is determined by the kinematic coverage of
the data (see Fig.~\ref{fig:kinrange}), as reflected by the errors on
the moments. In particular, at low $Q^2$ there are no large $x$ data,
while at high $Q^2$ there are no small $x$ data. In view of the fact
that the use of truncated moments allows one to exclude the small $x$
region, and that at low $Q^2$ power corrections can be sizable,
it is convenient to consider only the large $Q^2$ region. Also,
the large $x$ extrapolation is severely constrained by the kinematic
bound
$F_2(1,Q^2)=0$, whereas the  small $x$ behaviour is essentially
unknown~\cite{smallx}.
 A reasonable
cut which ensures a good coverage of the large $x$ region is $Q^2 \ge
20 $~GeV$^2$.  By choosing a high enough truncation point, $x_0 \gsim
0.3$, it would be possible to compute accurately truncated moments up
to the highest available scale $Q^2 \approx 200 $~GeV$^2$. Such a
choice, however, is not advisable, because the bulk of the data would
then be excluded. In fact, as $x_0$ is raised, correlations between
truncated moments rapidly increase. The value of $x_0$ should thus be
chosen as small as possible, in order not to loose information.
Requiring $x_0$ to be as low as $x_0 = 0.03$ imposes then a cut $Q^2
\le 70 $~GeV$^2$. Smaller values of $x_0$ would require extrapolation.

We take thus as a baseline choice $x_0 = 0.03$, $20 \le Q^2 \le 70
$~GeV$^2$. With this choice, moments $1<n <8$ have errors below $10\%$, and
correlation coefficients are below $0.9$, as long as neighbouring
moments are avoided, and one does not consider more than three
(logarithmically) equally spaced scales in the available $Q^2$
range. Correlations between moments are not significantly reduced by
further lowering $x_0$, which would anyway require extrapolation. If,
however, $x_0$ is raised to $x_0 = 0.1$, then correlations between
neighbouring moments are greater than $0.98$, and only correlations
between moments differing by more than two orders are below
$0.9$. Because correlations between the same moment evaluated at
different scales cannot be reduced without enlarging the $Q^2$ range,
no more than three scales should be used if we wish to keep such
correlations below $0.9$.  We will later consider variations of $x_0$,
the $Q^2$ range and the number of scales about this choice.

\subsection{Evolution equation}
\label{choev}

As discussed in \secn{trumo}, the scale dependence of any truncated
moment $q_n(x_0, Q^2)$ can be determined to any required accuracy from
the knowledge of a finite set of truncated moments at a reference
scale $Q_0^2$. The result has the form
\beq
q^{th}_n (x_0, Q_i^2) \equiv \sum_{p = n_{min}}^{M} M_{n p}(x_0; 
Q^2_0, Q^2_i; \as) ~q_p(x_0, Q_0^2)~,
\label{evolsol}
\eeq
where the evolution matrix $M_{np}$, explicitly given in
Refs.~\cite{us3,deer}, is determined as a perturbative expansion in
$\as$.

Given a measurement of truncated moments $q^{exp}_n(x_0,Q^2)$ at more
than one scale, the value of $\as$ can be determined by minimizing
\beq
\chi^2 = \sum_{n,i} \sum_{m,j} 
\left[ q^{exp}_n(x_0, Q_i^2) - q^{th}_n(x_0, Q_i^2) \right] 
V^{-1}_{n i; m j}
\left[ q^{exp}_m(x_0, Q_j^2) - q^{th}_m(x_0, Q_j^2) \right]~,
\label{chisq}
\eeq
where $V_{n i; m j}$ is the covariance matrix for moments $q^{exp}_n
(x_0, Q_i^2)$, $q^{exp}_m (x_0, Q_j^2)$.  Of course, the result for
$\as$ should be independent of the choice of initial scale $Q_0^2$.

If knowledge of moments $q_n (x_0, Q_0^2)$ with $n_{min} \le n \le M$
is needed in order to obtain the desired accuracy in the solution of
the evolution equation, then in principle all these moments should be
treated as free parameters in the minimization, along with the value
of $\as$, so the sum over $n, m$ in \eq{chisq}\ should run from
$n_{min}$ to $M$. In practice, however, for any reasonable value of
$x_0$ (say, $x_0 \lsim 0.1$), the evolution of truncated moment $q_n
(x_0, Q_0^2)$ is ``almost diagonal'', in the sense that it receives
only a small correction from moments $q_m (x_0, Q_0^2)$, with $m \not=
n$. Because of this, and because of the large correlations between
different moments, we may  perform the minimization while only
including a subset of moments in the sum over $n, m$ in \eq{chisq},
and treating only such moments as free parameters. This issue is
discussed in detail in the next subsection.

Throughout this section, evolution will be performed using the
improved method of Ref.~\cite{deer}, discussed in section
2.3. Specifically, we will take $n_{min} = 1$, with twelve moments
included in the evolution equation ($M = 11$), while the
reconstruction of $q(x_0, Q_0^2)$ according to \eq{pdfapprox} will be
performed with $N = 6$.  This ensures an accuracy of order $0.1\%$ on
the evolution of the second ($n = 2$) truncated moment, rapidly
improving as $n$ increases. For $\as$, we use the solution of the
next-to-leading order renormalization group equation to express
$\as(Q^2)$ in terms of $\as(M_Z)$, which is then directly taken as a
free parameter. The number of active flavors varies by one unit at
each quark threshold, and the coefficients of the $\beta$ function are
matched according to the Marciano~\cite{marciano} prescription.
Dependence on quark thresholds will be studied as a source of
theoretical uncertainty in \secn{therr}.

\subsection{Choice of fitted moments}
\label{chomo}

Having fixed the number of truncated moments to be included in the
evolution equation, one still has to choose which ones should be
treated as free parameters in the minimization procedure, and which
ones should be fixed at the experimental central value.
\begin{table}[t]  
\begin{center}  
\begin{tabular}{|l|c|c|c|} \hline 
$n$ & $x_0=0.1$ & $x_0=0.03$ & $x_0=0.01$ \\
\hline  
2  & 
     0.091 $\pm$ 0.047 & 0.085  $\pm$  0.070 & 0.089 $\pm$ 0.080 \\
3  & 
     0.100 $\pm$ 0.024 & 0.106  $\pm$  0.030 & 0.106 $\pm$ 0.031 \\
4  & 
     0.113 $\pm$ 0.019 & 0.115  $\pm$  0.019 & 0.115 $\pm$ 0.019 \\
5  & 
     0.122 $\pm$ 0.015 & 0.123  $\pm$  0.015 & 0.123 $\pm$ 0.015 \\
6  & 
     0.127 $\pm$ 0.014 & 0.127  $\pm$  0.014 & 0.127 $\pm$ 0.014 \\
7  & 
     0.129 $\pm$ 0.015 & 0.129  $\pm$  0.014 & 0.129 $\pm$ 0.015 \\
8  & 
     0.129 $\pm$ 0.016 & 0.129  $\pm$  0.016 & 0.129 $\pm$ 0.016 \\
9  & 
     0.129 $\pm$ 0.018 & 0.129  $\pm$  0.018 & 0.129 $\pm$ 0.018 \\
\hline
\end{tabular}
\end{center}
\caption{}{Fits of $\as(M_Z)$ from the evolution of a single moment.}
\label{assinglemom}
\end{table}
The simplest possibility is to fit only  one moment at a time.
 The results for $\as$ obtained in this case are
displayed in Table~\ref{assinglemom}, with the  default choice of
scales, and three different choices of truncation point $x_0$.  The
table shows that the uncertainty in the determination of $\as$, as a
function of the order of the moment $n$, has a minimum around $n
\approx 6$. The presence of a minimum can be understood as a
consequence of the fact that the most accurate data have large $x
\approx 0.5$, but there are no data with $x > 0.75$; furthermore, the
anomalous dimension vanishes at low $n \sim 1$ (it vanishes exactly at
$n = 1$ when $x_0 = 0$), and increases monotonically in modulus as $n$
increases. Hence, low moments lead to a less precise determination
both because they are dominated by small $x$ data and because their
scaling violations are weaker; for high enough $n$, on the other hand,
there is loss of precision due to the extrapolation in the very large
$x$ region. In particular (see also below) moments with $n > 9$ probe
a region of $x$ where elastic contributions to the cross-section
start being relevant, and become rapidly unreliable as $n$ increases.

It is interesting to observe that all the determinations of
Table~\ref{assinglemom} are compatible within errors. Also, it is
interesting to observe that, for the values of $n$ which give a
reasonably precise determination of $\as$, there is little or no
dependence on the choice of $x_0$ of both the central value and 
the error. The error on any of these individual determinations is
however much larger than the error obtained from existing fits to
these data~\cite{BCDal,NMCal}: a more accurate determination can only
be obtained by combining the information from different truncated
moments, \ie\ fitting more than one truncated moment at a time.
\begin{table}[t]  
\begin{center}  
\begin{tabular}{|l|c|} \hline 
\multicolumn{2}{|c|}{$x_0=0.03$}\\ \hline  
Fitted moments  & $\as$  \\
\hline
3+4  &  0.137  $\pm$ 0.011 \\
2+4  &  0.140  $\pm$ 0.010 \\
3+5  &  0.136  $\pm$ 0.011 \\
4+6  &  0.131  $\pm$ 0.012 \\
5+7  &  0.128  $\pm$ 0.012 \\
\hline
\end{tabular}
\end{center}
\begin{center}  
\caption{}{Fits of $\as(M_Z)$ from the evolution of a pair of moments.}
\label{aspairmom}
\end{center}  
\end{table}

When including more than one moment in the sum of \eq{chisq}, the
issue of correlations between moments becomes important.  The impact
of correlations is illustrated by the results obtained fitting a pair
of moments, displayed in Table~\ref{aspairmom}. Because of large
correlations, in each case the value of $\as$ turns out to be larger
than either of the values obtained from each of the two moments
individually.

The presence of large correlations entails two distinct problems when
performing a fit where several moments are simultaneously fitted. The
first problem is that, as the elements of the covariance matrix $V_{i
j} \to 1$, the matrix becomes singular, in that all eigenvalues but
one vanish in the limit. Hence, when correlations are large, several
eigenvalues become very small and the inversion of the covariance
matrix is numerically problematic.
\begin{figure}[t]
\begin{center}
\epsfig{width=0.7\textwidth,figure=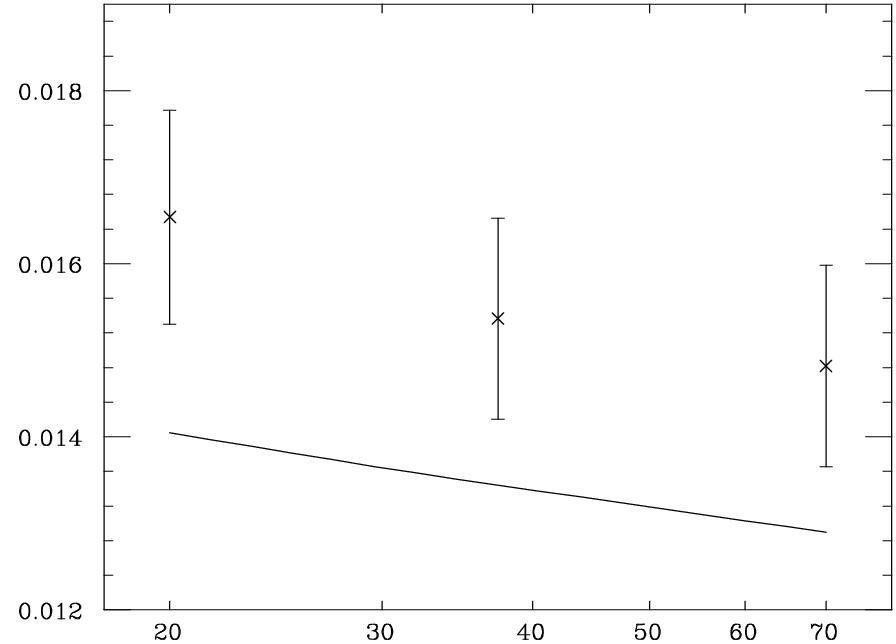}  
\end{center}
\caption{}{Pathological best fit of the third moment in the presence
of off-diagonal instabilities.}
\label{fig:pathol}
\end{figure}
The second problem is that, even if the covariance matrix is
accurately inverted, when correlations are large, off-diagonal terms
in the $\chi^2$ \eq{chisq} may dominate over diagonal ones, thereby
leading to generally unreliable and often pathological results. In
particular, it may turn out that the best-fit values of all moments
differ from the measured values by several standard deviations. An
example of this pathological situation is displayed in
Fig.~\ref{fig:pathol}, which shows the best-fit behaviour of the
third moment ($n = 3$) in a fit of moments $2 + 3 + 4 + 5 + 6$, chosen
as a representative case in which moments are
over-correlated because too many neighbouring moments are 
fitted.\footnote{Note that even though this pathological
situation is reminiscent of the effect which is found when linearizing
correlated normalization errors~\cite{dago}, it is a distinct
problem. To check this, we have replaced in the $\chi^2$ \eq{chisq}
all $q_n$ with $\ln q_n$, and computed the covariance matrix
accordingly. In such case normalization uncertainties decouple, yet we
have verified that off-diagonal instabilities are still present.}
Even though this situation is in principle acceptable, in practice
such results are unreliable because the minimum of the $\chi^2$ is
obtained by a fine-tuned balance of diagonal and off-diagonal
contributions, which cannot be trusted whenever errors and
correlations are known with limited accuracy~\cite{cowan}. We will
refer to this situation as {\it off-diagonal instability} of the
fits. 

Because of these problems, there is a trade-off in accuracy when
including more than one moment in the fit: the addition of more
moments brings in new information, thus reducing statistical errors,
but for a large enough number of moments it is impossible to
accurately invert the covariance matrix, and fits are spoiled by
instabilities. Whereas it is possible to keep the covariance matrix
inversion under control by improving the numerical accuracy of
calculations, the off-diagonal instability depends on the accuracy in
the experimental determination of errors and correlations, and it is
unavoidable. We have thus made sure that the covariance matrix is
inverted to an accuracy which is by several orders of magnitude
greater than the knowledge of its individual matrix elements. Then, we
have tested for off-diagonal instabilities by flagging all results in
which, at the best fit, one or more moments differ by more than one
standard deviation from the experimental central value at any of the
fitted scale, and discarding fits for which this happens for more than
one experimental point.

Clearly, the maximum number of moments which may be included in the
fit before an off-diagonal instability appears will be larger when
correlations are lower. This means that first, it is not advantageous
to further increase the number of scales beyond three (compare with
Table~\ref{corscal}); second, it is convenient not to fit
simultaneously neighbouring moments (compare with Table~\ref{cormom});
finally, it is convenient to choose a value of $x_0$ which is as low
as possible since, as discussed above, correlations rapidly increase
with $x_0$. Indeed, we find that, with three scales and $x_0 = 0.03$,
fits of up to five moments are possible, while with $x_0 = 0.1$ fits
of at most four moments are possible without incurring in
instabilities. With $x_0 = 0.01$ fits of up to seven moments are
possible, but this choice of $x_0$ requires a considerable amount of
extrapolation.
\begin{table}[t]  
\begin{center}  
\begin{tabular}{|l|c|} \hline 
\multicolumn{2}{|c|}{$x_0=0.03$}\\ \hline  
Fitted moments  & $\as$  \\
\hline
2+3+4    &   0.126 $\pm$ 0.010  \\
2+4+6    &   0.140 $\pm$ 0.008  \\
3+5+7    &   0.138 $\pm$ 0.009  \\
\hline
2+4+6+8  &   0.142 $\pm$ 0.009  \\
3+5+7+9  &   0.124 $\pm$ 0.007  \\
2+4+5+7  &   0.141     $\pm$ 0.009                \\
\hline
3+4+5+6+7  & 0.1256 $\pm$ 0.0049  \\
3+4+5+6+8 & 0.1247  $\pm$ 0.0050 \\
2+4+5+6+8 & 0.1242  $\pm$ 0.0042   \\
2+4+5+7+8 & 0.1254  $\pm$ 0.0044 \\
\hline
\end{tabular}
\end{center}
\caption{}{Fits of $\as(M_Z)$ from the evolution of an increasing number
of moments, with optimal $x_0$.}
\label{asmoremom}
\end{table}

The results for representative fits with $x_0=0.03$
are shown in Table~\ref{asmoremom}  
as an increasing number of moments
is fitted. It
is clear that both the size of the error and the stability of the
central value improve as the number of fitted moments
increases. Stability of the central value of $\alpha_s(M_z)$ 
is found with the largest number of fitted
moments allowed  before the onset of off-diagonal instabilities.

\subsection{Variation of the fit parameters}
\label{chosc}

We now turn to the effect of variations of the truncation point $x_0$
and of the range and number of scales.  Comparing the results of
Table~\ref{asmoremom} with those obtained with a larger value of $x_0$
shows that with higher $x_0$ it is not possible to achieve
satisfactory stability of the best-fit value of
$\as$ before the onset of off-diagonal
instabilities.  Some representative results obtained with $x_0 = 0.1$
are displayed in Table~\ref{as01}; similar results are obtained for
other values $x_0 > 0.03$. If instead $x_0$ is lowered to $x_0 =
0.01$ or below, the error  on the best-fit $\as$ does not improve further,
despite the fact that fits with a larger number of moments are
possible, and remains in fact somewhat larger than the best fit of
Table~\ref{asmoremom}. This is consistent with the fact that lowering
$x_0$ below $0.03$ does not introduce any new experimental
information.
\begin{table}[t]  
\begin{center}  
\begin{tabular}{|l|c|} \hline 
\multicolumn{2}{|c|}{$x_0=0.1$}\\ \hline  
Fitted moments  & $\as$  \\
\hline
2+3+4    &  0.128 $\pm$  0.008 \\
2+4+6    &  0.120 $\pm$  0.010 \\
3+5+7    &  0.121 $\pm$  0.015 \\
\hline
2+4+6+8  &  0.137 $\pm$  0.008 \\
3+5+7+9  &  0.140 $\pm$  0.012 \\
2+4+5+7  &  0.114 $\pm$  0.009 \\
\hline
\end{tabular}
\end{center}
\caption{}{Fit of $\as(M_Z)$ from the evolution of an increasing 
number of moments with large $x_0$.}
\label{as01}
\end{table}

Coming now to scale choices, the first issue is the dependence of our
results on the reference scale $Q_0^2$.  If we were fitting all
moments, results would be entirely independent of this scale, except
insofar as different choices of $Q_0^2$ correspond to different
choices for the first guess of the values of the moments in the
minimization routine. However, since only a subset of moments is
fitted, there might be a residual dependence on $Q_0^2$, due to the
fact that the scale dependence of the central values of the moments
which are not fitted might not agree completely with the predicted
scale dependence.  We have found that, in fact, the choice of initial
scale may affect the onset of off-diagonal instabilities: for
instance, the fit of moments $2 + 4 + 5 + 6 + 8$ turns unstable if
$Q_0^2 > 40 $~GeV$^2$. This is not surprising, in view of the fact
that the central values of low moments are less accurate at this
scale, because of the need to extrapolate at small $x$. Indeed, the
instability does not appear if evolution is performed using the method
of Ref.~\cite{us1}, instead of that of Ref.~\cite{deer}: that method
is less accurate, but it does not require knowledge of low moments since
the evolution matrix is triangular (as discussed in
\secn{trumo}). Nevertheless,  we
find that, in all cases in which a stable fit is obtained, the results
turn out to be essentially independent of the choice of the reference
scale $Q_0^2$.
\begin{table}[t]  
\begin{center}  
\begin{tabular}{|l|c|} 
\hline  
$Q^2$ range (GeV$^2$)& $\as$  \\
\hline
20-70    & 0.1242 $\pm$ 0.042 \\
20-100   & 0.1239 $\pm$ 0.049 \\
30-70   & 0.1239 $\pm$ 0.052 \\
30-100   & 0.1249 $\pm$ 0.059 \\
\hline
\end{tabular}
\end{center}
\caption{}{Dependence of the value of $\as$ on the $Q^2$ range for 
the $2 + 4 + 5 + 6 + 8$ fit with $x_0 = 0.03$}
\label{asq2}
\end{table}

Next, we consider the dependence on the choice of fitting scales.  The
range of scales cannot be widened much without requiring considerable
extrapolation. The effect of small variations is displayed in
Table~\ref{asq2}. It is apparent that no significant dependence is
found, and in fact the smallest error is obtained when $20 \leq Q^2 \leq
70 $~GeV$^2$. If instead the number of scales is varied, we find that
with only two scales the quality of the fit deteriorates considerably:
for instance, the error on $\as$ from the fit of moments $2 + 4 + 5 +
6 + 8$ with $x_0 = 0.03$ goes up to $\sigma = 0.0077$ if only two
scales are used. If four or more scales are used, fits with five
moments become unstable because of excessive correlations, and it
becomes impossible to find a stability region. Hence the choice of
three scales in the range $20 \leq Q^2 \leq 70 $~GeV$^2$ appears to be
optimal.

\subsection{Best fit}
\label{befit}

\begin{figure}[t]
\begin{center}
\epsfig{width=0.7\textwidth,figure=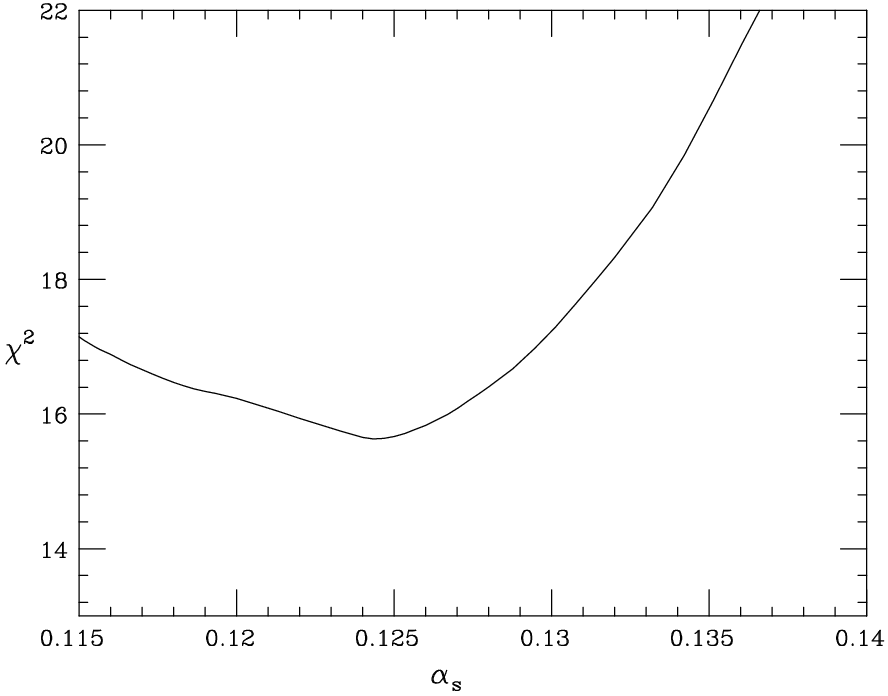}  
\end{center}
\caption{}{$\chi^2$ as a function of $\as(M_Z)$ for the fit of 
moments $2 + 4 + 5 + 6 + 8$ with $x_0 = 0.03$.}
\label{fig:scanning}
\end{figure}

Having tested how the quality of the fit varies with all the choices
enumerated in the previous subsections, we select the fit architecture
that maximizes the stability and minimizes the error.  Our conclusion
is that a reliable and stable determination of $\as(M_Z)$ is obtained
with $x_0 = 0.03$, five moments (Table~\ref{asmoremom}) and three
scales $20 \leq Q^2 \leq 70 $~GeV$^2$. Specifically, the smallest
error is obtained with the `symmetric' combination of moments $2 + 4 +
5 + 6 + 8$, which we take as our baseline result. Note that the
central value of $\as$ coincides with that which is obtained from
single-moment fits (Table~\ref{assinglemom}) when the error is
stationary, \ie\ for $5 \leq n \leq 6$.  In order to obtain a more
accurate determination of the statistical error, we have studied the
dependence of the $\chi^2$ on the value of $\as$, displayed in
Fig.~\ref{fig:scanning}. The $\chi^2$ is asymmetric about the minimum,
rising more slowly as $\as$ decreases.  We arrive thus at the
determination
\beq
\as(M_Z) = 0.124 ~\epm{0.004}{0.007} ~\hbox{(stat.)}~.
\label{alstat}
\eeq

\subsection{Theoretical uncertainties}
\label{therr}

The main sources of theoretical uncertainty are higher-order
perturbative and higher-twist corrections, as well as heavy quark
threshold effects. First of all, our fits are based on leading-twist
evolution of structure functions, so one should worry about possible
power corrections. The largest corrections of this kind are target
mass corrections, which are known analytically~\cite{tmc}. We have
checked that these corrections are less than $1\%$ on any of the
moments included in our fits. Because higher-twist corrections to the
operator product expansion are known~\cite{ht} to be significantly
smaller than target-mass corrections, we conclude that all power
corrections are entirely negligible in our analysis, thanks to the
relatively high cut $Q^2 \ge 20 $~GeV$^2$ which we have imposed. Also,
one might worry that very high moments might be sensitive to elastic
contributions to the cross section~\cite{elastic}.  We have verified
that such contributions are below $1\%$ for all moments if $Q^2 \ge 30
$~GeV$^2$, and reach at most $3\%$ for the $8$-th moment at $Q^2 = 20
$~GeV$^2$, which is the highest fitted moment. We conclude that these
contributions are also negligible. Nuclear corrections to
the deuterium structure function affect
the initial values of the fitted moments but not their  scale
dependence, and are thus immaterial, up to  nuclear
higher twist corrections, which are not expected to be significantly
larger than standard higher twist terms (except possibly at very small
$x$)~\cite{nucl}. 

We are then left with uncertainties related to higher-order
perturbative corrections and to heavy quark thresholds. The position
of thresholds $Q^2 = k_{th} M_q^2$ has been varied in the range $0.3
\le k_{th} \le 4$. Higher-order corrections have been estimated by
varying the renormalization scale $\mu^2_{ren} = k_{ren} Q^2$ in the
standard way~\cite{scalvar}, with $0.3 \le k_{ren} \le 4$. Notice
that, because the structure function is evolved directly in  the DIS
scheme~\cite{DIS}, 
there is no factorization scale dependence.  The
$\chi^2$ of the fit is almost independent of the choice of $k_{th}$,
while it tends to increase considerably if $k_{ren} < 0.5$ or $k_{ren}
> 2$; in particular, with $k_{ren} \le 0.25$ we were unable to obtain
a stable
fit. 

\begin{figure}[t]
\begin{center}
\epsfig{width=0.45\textwidth,figure=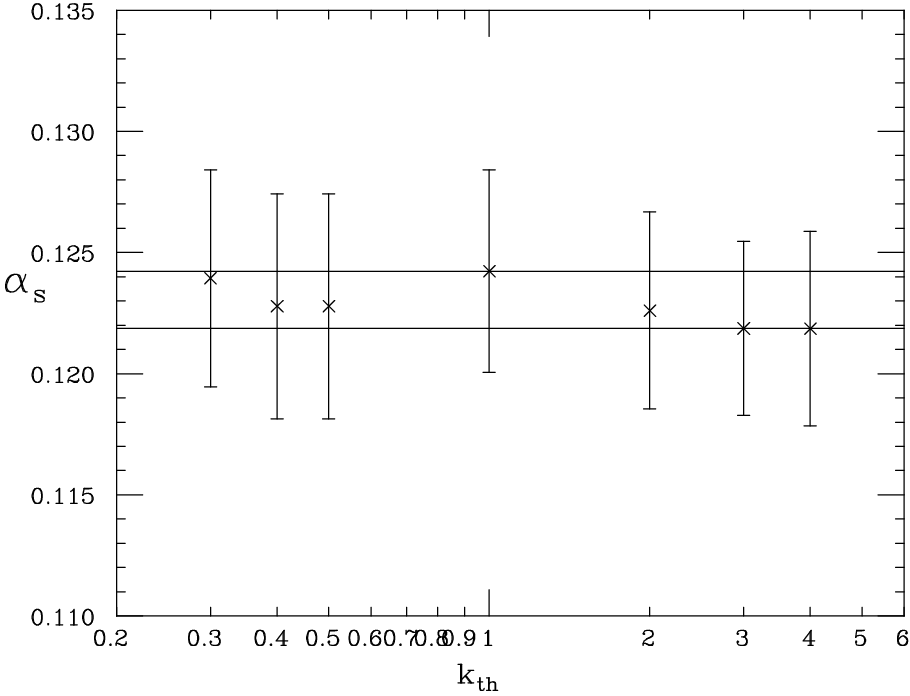}  
\epsfig{width=0.45\textwidth,figure=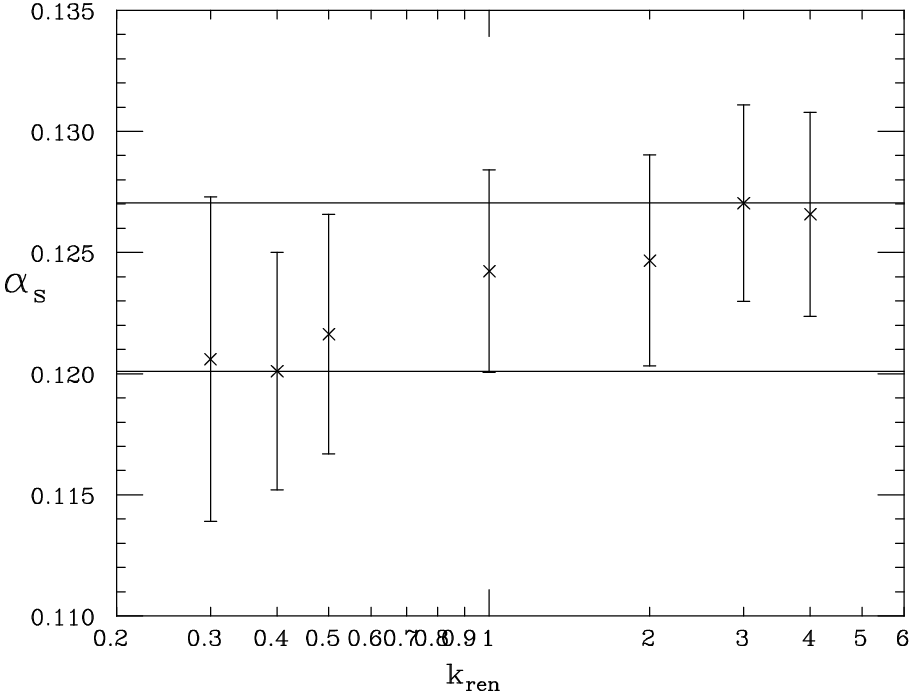}  
\end{center}
\caption{}{Dependence of the best-fit value of $\as(M_Z)$ on the
position of heavy quark thresholds (left) and renormalization scale
(right). The band indicates the overall uncertainty.}
\label{fig:syst}
\end{figure}

The dependence of the value of $\as(M_Z)$ on $k_{th}$ and
$k_{ren}$ for the fit of moments $2 + 4 + 5 + 6 + 8$ with $x_0 = 0.03$
and three scales is displayed in Figure~\ref{fig:syst}. The associated
uncertainties are estimated to be
\beq
\sigma(\hbox{thresh.}) = \epm{0.000}{0.002}~; \qquad
\sigma(\hbox{ren.})    = \epm{0.003}{0.004}~.
\label{syserr}
\eeq
The dependence on the position of thresholds is, predictably, very
weak, given that the $b$ threshold is close to the edge of our $Q^2$
range, and falls outside it as soon as $k_{th} < 0.8$.  The dependence
on renormalization scale turns out to be also reasonably weak.

\subsection{Result and comments}
\label{resco}

Our final determination of the strong coupling is
\beq
\as(M_Z) = 0.124 ~\epm{0.004}{0.007}~\hbox{(exp.)} ~\epm{0.003}{0.004} 
{} ~\hbox{(th.)} = 0.124 ~\epm{0.005}{0.008} ~\hbox{(total)}~. 
\label{alfin} 
\eeq
The error on the result is dominated by statistical uncertainties,
consistent with our expectation that the method of analysis used here
minimizes theoretical uncertainties.

The value of $\as$ has been previously extracted from QCD fits to the
BCDMS data, with the result~\cite{BCDal} $\as(M_Z) = 0.113 \pm 0.003
\hbox{(exp.)} \pm 0.004 \hbox{(th.)}$, and to the NMC data, with the
result~\cite{NMCal} $\as(M_Z) = 0.117 \epm{0.011}{0.016}$. There are
two main differences between these previous determinations and the
present one, \eq{alfin}. First, the determinations in
Refs.~\cite{BCDal,NMCal} were based on global QCD fits, and thus
required the construction of a parton parametrization, which, as
discussed in the introduction, might be the source of systematic error
and bias. Second, they did not include a full treatment of correlated
systematics: statistical and systematic errors were added in
quadrature. A reanalysis of the BCDMS data which did include a
treatment of correlated systematics found instead~\cite{BCDre}
$\as(M_Z) = 0.118 \pm 0.002~\hbox{(exp.)}$. Our value appears thus in
good agreement with other determinations from the same data.  While a
direct comparison of the uncertainties is difficult, we find it
interesting that, while the overall uncertainty of our value is very
close to that of the BCDMS determination~\cite{BCDal}, our analysis
gives excellent control on theoretical uncertainties: the dominant
error is experimental and could be improved upon by future
experiments.

It is interesting to observe that the central value we find, though in
agreement within error with  current global averages, is on the high side. As it appears
from Table~\ref{assinglemom}, the central value of $\as$ tends to be
higher for higher moments. This is very suggestive of soft gluon
resummation effects: as is well known~\cite{sudakov}, leading higher
order corrections at large $N$ are resummed by the replacement $Q^2
\to Q^2/N$ in the argument of $\as$ in the evolution equation for the
$N$-th (full) moment. This means that the effective value of $\as$ is
larger for the evolution of higher moments. Our results provide some
indication of this effect, and suggest that a better determination of
$\as$ could be obtained from these data if this resummation were
included, by generalizing the soft gluon resummation formalism to the
case of truncated moments. In particular, the inclusion of soft gluon 
effects to all orders is likely to reduce the theoretical uncertainty
expressed by the dependence on the renormalization scale.

\section{Concluding remarks}
\label{conre}

We have presented a determination of $\alpha_s$ from scaling
violations aimed at minimizing the sources of theoretical bias
which might be cause of concern in existing determinations of
$\alpha_s$ from deep inelastic scattering experiments.  This has been
accomplished by avoiding the use of a parton parametrization: thanks
to the use of truncated moments, we have directly fitted the scaling
violations of a physical observable. Truncated moments in turn have
been determined  by means of a bias-free parametrization
of the structure function, inferred from the data, which retains all
the information on experimental errors and correlations.

It is interesting to compare our final result for $\alpha_s$ to other
determinations from deep inelastic scattering, and in particular those
obtained using the same data set~\cite{BCDal,NMCal,BCDre}. Whereas in our
determination theoretical errors are small and fully under control,
the price to pay for this is that the experimental error is larger
than in other determinations. It is unclear to which extent this is a
trade-off, or a consequence of the need to impose restrictive cuts in
the $Q^2$ range in order to deal with truncated moments. However, it
does suggest that a substantially more precise determination could be
obtained using data with either smaller statistical error (especially
for deuteron data), or spanning a wider $Q^2$ range, or both, such as could be
obtained for instance at the planned EIC facility~\cite{eic}.

As far as the central value of $\alpha_s$ is concerned, it is
suggestive that our determination is on the high side, since this is
what one would expect in the presence of sizable soft gluon
resummation effects. These in turn are expected to be more important
in our determination, since the kinematic cuts which we imposed give
more weight to the large-$x$ region, where such effects are
larger. Therefore, we expect that a resummed version of our analysis
might lead to a more accurate result for $\alpha_s$, and provide
explicit evidence for soft gluon resummation.

Finally, our analysis provides an explicit demonstration of the power
of methods of analysis based on the direct determination of the
probability density of physical observables from the data, and their
use coupled with the bias-free computational method which is afforded
by the use of truncated moments.

\vskip 1cm

{\large {\bf Acknowledgements}} 

\vskip 3mm

It is a pleasure to thank L. Garrido, who co-authored Ref.~\cite{us2}
and participated in many  discussions along the course of this project.  We also
thank G.~Altarelli, G.~d'Agostini and G.~Ridolfi for discussions.
This work was supported in part
by EU TMR contract FMRX-CT98-0194 (DG 12-MIHT) and by the Spanish
and Catalan grants AEN99-0766, AEN99-0483, 1999SGR-00097.

\vfill
\eject


\begin{thebibliography}{99}
 
\baselineskip14pt
 
\bibitem{qcd} 
See {\it e.g.} S.~Catani {\it et al.}, in {\it CERN 1999, 
Proceedings} ``Standard Model Physics (and more) at the LHC'', 
{\it CERN report} {\bf CERN-2000-004}, {\tt hep-ph/0005025};\\
S.~Alekhin {\it et al.}, {\tt hep-ph/0204316}, to be published 
in {\it Les Houches 2001, Proceedings} ``Physics at TeV Colliders''.

\bibitem{sb} 
S.~Bethke, {\it J. Phys.} {\bf G26} (2000) R27,
{\tt hep-ex/0004021}.

\bibitem{altarev}
G.~Altarelli,
{\tt hep-ph/0204179}.

\bibitem{AP} 
G.~Altarelli and G.~Parisi, \np{B126}{77}{298}.

\bibitem{esw} 
See {\it e.g.} R.K.~Ellis, W.J.~Stirling and B.R.~Webber, 
``QCD and Collider Physics'' (Cambridge, 1996).

\bibitem{pdfrev} 
See {\it e.g.} S.~Forte, {\it Nucl. Phys.} {\bf A666} (2000) 113.

\bibitem{abfr} 
G.~Altarelli {\it et al.}, {\it Nucl. Phys.} {\bf B496}, 337 (1997), 
{\tt hep-ph/9701289}; 
{\em Acta Phys. Pol.} {\bf B29} (1998) 1145, {\tt hep-ph/9803237}.

\bibitem{pdfer} 
W.T.~Giele, S.A.~Keller and D.A.~Kosower, {\tt hep-ph/0104052};
also in {\it La Thuile 1999, Proceedings} ``Rencontres de
Physique de la Vall\'ee d'Aoste'', p.~255 (INFN, Frascati, 1999);
W.~T.~Giele and S.~Keller, \pr{D58}{98}{094023}, {\tt hep-ph/9803393}.

\bibitem{us1} 
S.~Forte and L.~Magnea, \pl{B448}{99}{295}, {\tt hep-ph/9812479};
also in {\it Tampere 1999, Proceedings} ``International Europhysics 
Conference on High-Energy Physics'' (EPS--HEP 99), Tampere 1999, p.~454, 
{\tt hep-ph/9910421}.

\bibitem{us3} 
S.~Forte, L.~Magnea, A.~Piccione and G.~Ridolfi, {\it Nucl. Phys.} 
{\bf B 594} (2001) 46, {\tt hep-ph/0006273}.

\bibitem{deer} 
A.~Piccione, {\it Phys. Lett.} {\bf B518} (2001) 207, {\tt hep-ph/0107108}.

\bibitem{us2} 
S.~Forte, L.~Garrido, J.I.~Latorre and A.~Piccione, {\tt hep-ph/0204232}.

\bibitem{DIS}
G.~Altarelli, R.~K.~Ellis and G.~Martinelli, {\it Nucl.\ Phys.} 
{\bf B143} (1978) 521 [Erratum--ibid.\ {\bf B146} (1978) 544];
{\it Nucl.\ Phys.} {\bf B157} (1979) 461.

\bibitem{ale} 
S.I.~Alekhin, {\it Eur. Phys. J.} {\bf C10} (1999) 395,
{\tt hep-ph/9611213}; \\ 
D.~Stump {\it et al.}, {\it Phys. Rev.} {\bf D65} (2002) 014012,
{\tt hep-ph/0101051}; \\
J.~Pumplin {\it et al.}, {\it Phys. Rev.} {\bf D65} (2002) 014013,
{\tt hep-ph/0101032}.

\bibitem{ortho} 
F.J.~Yndurain, {\it Phys. Lett.} {\bf B74} (1978) 68; \\
G.~Parisi, N.~Sourlas, {\it Nucl. Phys.} {\bf B151} (1979) 421; \\
W.~Furmanski, R.~Petronzio, {\it Nucl. Phys.} {\bf B195} (1982) 237.

\bibitem{NMC} 
M.~Arneodo {\it et al.} (New Muon Coll.), {\it Nucl. Phys.} {\bf B483} 
(1997) 3, {\tt hep-ph/9610231}.

\bibitem{BCDMS} 
A.C.~Benvenuti {\it et al.} (BCDMS Coll.), {\it Phys. Lett.} {\bf B223} 
(1989) 485; {\it Phys. Lett.} {\bf B237} (1990) 592.

\bibitem{e665} 
M.R.~Adams {\it et al.} (E665 Coll.) {\it Phys. Rev.} {\bf D54} (1996) 3006.

\bibitem{netrev}
C. Peterson and T. R\"ognvaldsson, {\it Lectures at the 1991 CERN 
School of Computing, preprint} {\bf LU--TP--91--23}; \\
B. M\"uller, J. Reinhardt and M.~T.~Strickland, {\it Neural Networks: 
an introduction} (Berlin, 1995); \\
G.~Stimpfl--Abele and L.~Garrido, {\it Comput.\ Phys.\ Commun.}  
{\bf 64} (1991) 46.

\bibitem{smallx}
See {\it e.g.} S.~Forte and R.~D.~Ball,
{\it Acta Phys.\ Polon.} {\bf B26} (1995) 2097,
{\tt hep-ph/9512208}.

\bibitem{marciano} 
W.J.~Marciano, {\it Phys. Rev.} {\bf D29} (1984) 580.

\bibitem{BCDal} 
M.~Virchaux and A.~Milsztajn, {\it Phys. Lett.} {\bf B274} (1992) 221.

\bibitem{NMCal} 
M.~Arneodo {\it et al.}, {\it Phys. Lett.} {\bf B309} (1993) 222.

\bibitem{BCDre} 
S.I.~Alekhin, {\it Phys. Rev.} {\bf D59} (1999) 114016, {\tt hep-ph/9809544}.

\bibitem{cowan} 
G.~Cowan, {\it Statistical data analysis}, (Oxford, 1998).

\bibitem{dago}
G.~D'Agostini, {\it Nucl. Instrum. Meth.}  {\bf A346} (1994) 306.

\bibitem{tmc} 
See {\it e.g.} J.L.~Miramontes and J.~S\'anchez--Guillen, {\it Z. Phys.} 
{\bf C41} (1988) 247, and refs. therein.

\bibitem{ht} 
U.K.~Yang and A.~Bodek, {\it Phys. Rev. Lett.} {\bf 82} (1999) 2467, 
{\tt hep-ph/9809480}.

\bibitem{elastic} 
A.~Bodek, in {\it Blois 1994 Proceedings} ``The heart of the matter'', 
p.~255.

\bibitem{nucl}
See K.~J.~Eskola {\it et al.},
{\tt hep-ph/0110348} and ref. therein.

\bibitem{scalvar} See {\it e.g.}
G.~Ridolfi and S.~Forte, {\it J. Phys.} {\bf G25} (1999) 1555.

\bibitem{sudakov} 
See \eg\ S.~Catani, M.L.~Mangano, P.~Nason and L.~Trentadue, 
{\it Nucl. Phys.} {\bf B478} (1996) 273, {\tt hep-ph/9604351}, 
and refs. therein.

\bibitem{eic}
A.~L.~Deshpande, {\it Nucl.\ Phys.\ Proc.\ Suppl.} {\bf 105} (2002) 178.

\end{thebibliography}
\end{document}